\documentclass[prl,twocolumn,showpacs,superscriptaddress]{revtex4}
\usepackage{amsfonts}
\usepackage{amsmath}
\usepackage{dcolumn}
\usepackage{graphicx}
\usepackage{bm}

\DeclareMathOperator{\Tr}{Tr}

\DeclareMathOperator{\im}{Im}

\begin{document}

\title{Lattice Laughlin States of Bosons and Fermions at Filling Fractions $%
1/q$}

\author{Hong-Hao Tu}
\author{Anne E. B. Nielsen} 
\author{J. Ignacio Cirac}
\affiliation{Max-Planck-Institut f\"ur Quantenoptik, Hans-Kopfermann-Str.\ 1, D-85748 Garching, Germany}
\author{Germ{\'a}n Sierra}
\affiliation{Instituto de F\'isica Te\'orica, UAM-CSIC, Madrid, Spain}

\begin{abstract}
We introduce a two-parameter family of strongly-correlated wave functions for bosons and fermions in lattices. One parameter, $q$, is connected to the filling fraction. The other one, $\eta$, allows us to interpolate between the lattice limit ($\eta=1$) and the continuum limit ($\eta\to 0^+$) of families of states appearing in the context of the fractional quantum Hall effect or the Calogero-Sutherland model. We give evidence that the main physical properties along the interpolation remain the same. Finally, in the lattice limit, we derive parent Hamiltonians for those wave functions and in 1D, we determine part of the low energy spectrum.
\end{abstract}

\maketitle

\textit{Introduction.--} The fractional quantum Hall (FQH) effect has attracted a longstanding interest in physics. 2D electrons displaying such an effect form incompressible quantum liquids with a bulk gap, gapless edge states, and quasiparticle excitations with fractional charge and fractional statistics. Their properties are not amenable to the conventional Ginzburg-Landau theory; however, they can be thoroughly analyzed thanks to the discovery of analytical wave functions, which provide good approximations to some of the quantum states responsible for the FQH effect. An important family of such states is the Laughlin states \cite{Laughlin-1983}
\begin{equation}\label{eq:psiq}
\Psi_q (\{Z\})=\prod_{i<j}(Z_{i}-Z_{j})^{q}\exp \left(
-\sum_{l}|Z_{l}|^{2}/4\right),
\end{equation}
where $Z_{i}$ is the position in the complex plane of the $i$th electron and $\nu =1/q$ is the filling fraction, i.e., the ratio between the number of electrons and the number of flux quanta. From a modern viewpoint, the Laughlin states belong to the so-called topological phases \cite{Wen-1990,Wen-Niu-1990}, an exotic class of gapped phases whose full classification is still an outstanding open problem.

In the FQH setups, the Laughlin states arise due to the strong interactions
between the electrons in the fractionally filled lowest Landau level. In
that case, the size of the electron wave packets is at least one order of
magnitude larger than the lattice spacing and thus the lattice effects are
usually negligible \cite{JainBook}. A natural question is whether Laughlin states (or their variants) can appear in lattice models without Landau levels. In the late eighties, Kalmeyer and Laughlin (KL) proposed a state \cite{KL1987,KL1989,Laughlin-1989} that is a lattice version of the bosonic Laughlin state with $q=2$. This state has been shown to share some of the most defining properties of its continuum counterpart, like the fractional statistics of quasiparticle excitations \cite{KLCSL} and the presence of chiral edge states \cite{XGWen-1991}. Thus, the continuum and lattice version of the bosonic Laughlin state with $q=2$ seem to be closely connected, although it is not clear what such a connection is. In \cite{Scaffidi}, it has been shown that an interpolation Hamiltonian between a $q=2$ Laughlin-like lattice state and the continuum $q=2$ Laughlin state can be obtained by choosing bases that allow both states to be expressed in the same Hilbert space, although with different base kets. A more direct interpolation, in which the lattice spacing is continuously changed, has been considered in \cite{hafezi}, but was found to be valid only for sufficiently small lattice filling factors. A similar situation is encountered in 1D, where the Calogero-Sutherland (CS) model \cite{Calogero-1969,Sutherland-1971}, which is defined in the continuum, seems to be closely related to the Haldane-Shastry lattice model \cite{Haldane-1988,Shastry-1988}, although it is not obvious how to transform one into the other.

A very useful description of FQH wave functions in the continuum has been introduced by Moore and Read in \cite{Moore-Read-1991}, where they wrote selected FQH wave functions in terms of correlators of the corresponding edge conformal field theories (CFTs). Recently, for certain lattice systems in 1D and 2D, strongly correlated spin wave functions have also been written in terms of CFT correlators \cite{Ignacio-German-2010,nsc-2011,nsc-2012,Tu-2013}. This, in particular, has made it possible to construct parent Hamiltonians and to build in a systematic form simple wave functions with topological properties. We note also that parent Hamiltonians of the KL state have been found in \cite{Schroeter-2007,Thomale-2009,Kapit-2010,nsc-2012,Bauer-2013,nsc-2013}.

In this Letter, we provide an explicit connection between the continuum Laughlin/CS states on the one side and a set of lattice Laughlin/CS states on the other for all filling factors $1/q$. We do this by introducing a family of \emph{lattice} wave functions for hardcore bosons and fermions, which is defined on arbitrary lattices in 1D and 2D and allows us to continuously interpolate between the two limits. We also provide numerical evidence that the states remain within the same phase for all values of the interpolation parameter, so that the interpolation is meaningful. In 1D, we show that the states are critical and describe Tomonaga-Luttinger liquids (TLLs) with Luttinger parameter $K=1/q$, and in 2D we find that the states have topological entanglement entropy (TEE) $-\ln(q)/2$. The wave functions are constructed from conformal fields, and we use the CFT properties of the states to derive parent Hamiltonians for the wave functions in the lattice limit in both 1D and 2D and for general $q$. In 1D, the parent Hamiltonians are closely related to Haldane's inverse-square model \cite{Haldane-1988}, and we find that \emph{part} of the spectrum is given by integer eigenvalues described by a simple formula.

\textit{CFT wave functions.--} Let us consider a lattice with lattice
sites at the positions $z_{j}$, $j=1,2,\ldots ,N$, in the complex plane. The
local basis at site $j$ is labeled by $|n_{j}\rangle $, where $n_{j}\in
\{0,1\}$ is the number of particles at the site. The family of wave functions we propose (later on referred to as CFT states) take the form of the following chiral correlators of vertex operators:
\begin{equation}
\Psi (n_{1},\ldots ,n_{N})\propto\langle V_{n_{1}}(z_{1})\ldots
V_{n_{N}}(z_{N})\rangle ,  \label{eq:iMPS}
\end{equation}
where
\begin{equation}
V_{n_{j}}(z_{j})=\chi_j^{n_j}e^{i\pi \sum_{k(<j)} \eta_kn_{j}}:e^{i(qn_{j}-\eta_j )\phi (z_{j})/\sqrt{q}}:.  \label{eq:Vertex}
\end{equation}
Here, $\phi (z)$ is a chiral bosonic field from the $c=1$ free-boson CFT, $:\ldots :$ denotes normal ordering, $\chi_j$ are phase factors that do not depend on $n_j$, $q$ is a positive integer, and $\eta_j$ are positive parameters with average $N^{-1}\sum_j\eta_j=\eta\in(0,1]$. The charge neutrality condition $\sum_i(qn_i-\eta_i)=0$ of the CFT correlators fixes the number of particles to $\sum_{i=1}^{N}n_{i}=\eta N/q\equiv M$, which must hence be an integer, and it follows that $\eta /q$ is the lattice filling fraction. $\eta$ is therefore the parameter that interpolates between the continuum limit ($\eta\to0^+$), with infinitely many lattice sites per particle, and the lattice limit ($\eta=1$), in which the lattice filling fraction $\eta/q$ equals the Laughlin filling fraction $1/q$. When varying $\eta$, we shall always take all $\eta_j$ to scale linearly with $\eta$, such that $\eta_j/\eta_l$ remain constant. Evaluating the vacuum expectation value of the product of vertex operators in (\ref{eq:iMPS}) \cite{CFTbook} yields a Jastrow wave function
\begin{equation}
\Psi (n_{1},\ldots ,n_{N})\propto \delta
_{n}\prod_{i<j}(z_{i}-z_{j})^{qn_{i}n_{j}}\prod_{l}f_N(z_{l})^{n_{l}},
\label{eq:Laughlin}
\end{equation}
where $\delta _{n}=1$ if $\sum_{i=1}^{N}n_{i}=\eta N/q$ and zero otherwise
and $f_N(z_{l})\equiv \chi_{l}\prod_{j(\neq l)}(z_{l}-z_{j})^{-\eta_j} =\chi_{l}\exp[-\sum_{j(\neq l)}\eta_j\ln(z_{l}-z_{j})]$.

\begin{figure}[tbp]
\includegraphics[width=\columnwidth]{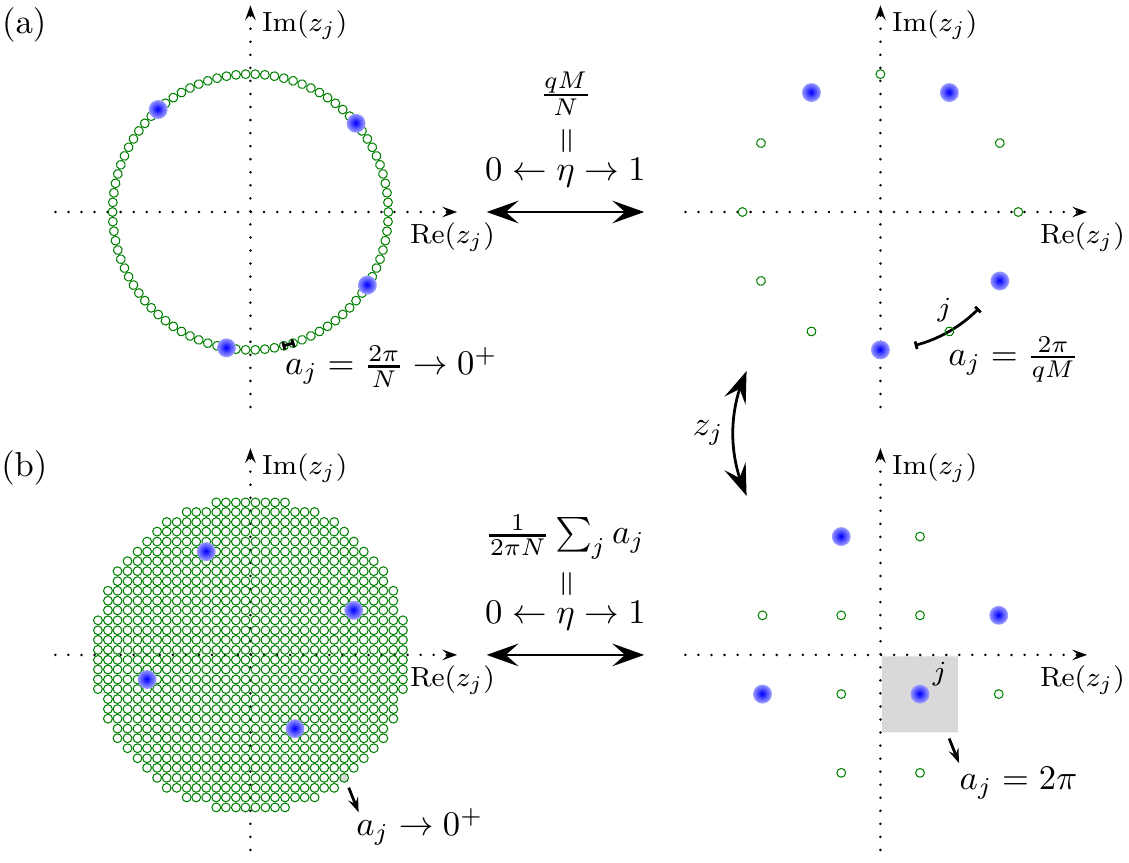}
\caption{(Color online) Illustration of the interpolation between the lattice limit ($\eta=1$) and the continuum limit ($\eta\to0^+$) for a uniform lattice in 1D and a square lattice in 2D. The interpolation is done, while keeping the area per particle $aN/M$ fixed, where $a$ is the average area per site. (a) In 1D, the lattice is defined by $z_j=e^{2\pi ij/N}$, which fixes the area of site $j$ to $a_j=2\pi/N$ $\forall j$, so that $a\equiv N^{-1}\sum_ja_j=2\pi/N$. The scaling parameter is therefore $\eta=qM/N=qMa/(2\pi)$. (b) In 2D, the lattice is defined on a disk with radius $R_\mathcal{D}\to\infty$, and we choose $a=2\pi qM/N$, since this fixes the area per particle to $2\pi q$ as in the Laughlin wave functions. The scaling parameter is therefore $\eta=qM/N=a/(2\pi)$. Transformations between different lattices, including the two displayed on the right, is obtained by transforming $z_j$.}
\label{fig:lattice}
\end{figure}

\textit{Relation to the CS and Laughlin wave functions.--} Let us demonstrate how the CFT states are related to several familiar wave functions in the continuum. We first consider the 1D periodic chain, where the lattice sites are uniformly distributed on a unit circle, i.e., $z_{j}=e^{2\pi ij/N}$, and we choose $\eta_j=\eta$ $\forall j$. In this case, we obtain analytically that $f_N(z_{l})\propto \chi_lz_{l}^\eta$, and we can therefore write the state (\ref{eq:Laughlin}) as a product of the wave function $\Psi_{\textrm{CS}}\propto \delta_{n}\prod_{i<j}(z_{i}-z_{j})^{qn_{i}n_{j}} \prod_{l}z_{l}^{-q(M-1)n_{l}/2}$ and the gauge factor
$\prod_{l}(\chi_l z_{l}^{\eta+q(M-1)/2})^{n_l}$. In the continuum limit, where $N\to\infty$, $\eta \to 0^{+}$, and $\eta N$ stays fixed to keep the number of particles $M$ and the area of the lattice constant (see Fig.~\ref{fig:lattice}(a)), the lattice spacing goes to zero, and $\Psi_{\textrm{CS}}$ turns into the ground-state wave function of the CS model \cite{Calogero-1969,Sutherland-1971} for bosons (even $q$) and fermions (odd $q$). The gauge factor can be set to unity by choosing $\chi_l=z_{l}^{-\eta-q(M-1)/2}$ if we like, but we note that its presence does not affect properties such as the particle-particle correlation function and the entanglement entropy. The CFT states thus allow us to define a lattice version of the CS wave functions and to interpolate between the lattice and the continuum limit of the model.

\begin{figure}
\includegraphics[width=\columnwidth]{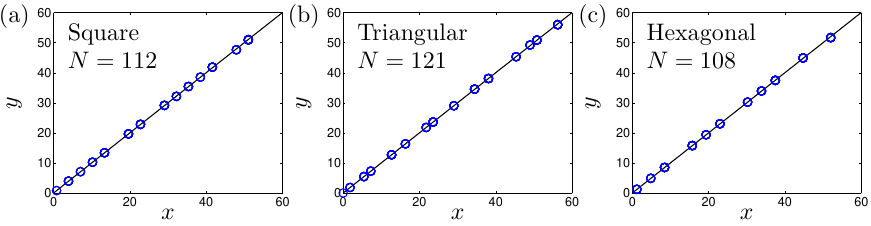}
\caption{(Color online) Numerical demonstration that \eqref{eq:fz} is approximately valid even for a moderate number of lattice sites $N$ for the square (a), the triangular (b), and the hexagonal (c) lattice with a circular edge. $x=|z_j|^2/4$, $y=-\ln[|f_N(z_j)|]+\textrm{constant}$, and the black lines in the background are the curve $y=x$.}\label{fig:fz}
\end{figure}

We next consider an arbitrary lattice in 2D, which is defined on a disk $\mathcal{D}$ of radius $R_\mathcal{D}\to\infty$. We define the area $a_j$ of site $j$ to be the area of the region consisting of all points in $\mathcal{D}$ that are closer to $z_j$ than to any of the other lattice sites. Let us note that $|f_N(z_{l})|=\exp[-\sum_{j(\neq l)}\eta_j\ln(|z_{l}-z_{j}|)]$. If we choose $\eta_j=a_j/(2\pi)$ and consider the continuum limit $\eta\to0^+$ (as illustrated for a square lattice in Fig.~\ref{fig:lattice}(b)), we can replace the sum over $j$ by the integral $\int_\mathcal{D}d^2z \ln(|z_l-z|)/(2\pi)$. In the thermodynamic limit $R_\mathcal{D}\to\infty$ this integral evaluates to $|z_l|^2/4+\textrm{constant}$, where the constant does not depend on $z_l$. Note, however, that $\sum_{j(\neq l)}\eta_j\ln(|z_{l}-z_{j}|)$ and $\kappa^{-2}\sum_{j(\neq l)}\kappa^2\eta_j\ln(|\kappa z_{l}-\kappa z_{j}|)$, where $\kappa$ is a positive constant, only differ by a $z_l$-independent constant for $R_\mathcal{D}\to\infty$. If $\eta_j$ is not small, we can choose $\kappa$ very small, transform the resulting sum into an integral, and again conclude that $\sum_{j(\neq l)}\eta_j\ln(|z_{l}-z_{j}|) =|z_l|^2/4+\textrm{constant}$. For all 2D lattices in the thermodynamic limit, we therefore obtain
\begin{equation}
f_N(z_l)\propto \chi_l e^{-ig_l}e^{-|z_l|^{2}/4} \quad (N\textrm{ large}), \label{eq:fz}
\end{equation}
where $g_l\equiv\im[\sum_{j(\neq l)}\eta_j\ln(z_{l}-z_{j})]$ is a real number. In Fig.~\ref{fig:fz}, we find numerically for different lattices that \eqref{eq:fz} is an accurate approximation even if $N$ is only moderately large. Choosing $\chi_l=e^{ig_l}$ and inserting \eqref{eq:fz} into \eqref{eq:Laughlin}, we observe that the CFT states coincide with the Laughlin states \eqref{eq:psiq}, except that the possible particle positions are restricted to the coordinates of the lattice sites. By changing the number of lattice sites per particle, we can thus interpolate between the Laughlin states in the continuum and Laughlin-like states on lattices.

\begin{figure}[tbp]
\includegraphics[width=\columnwidth]{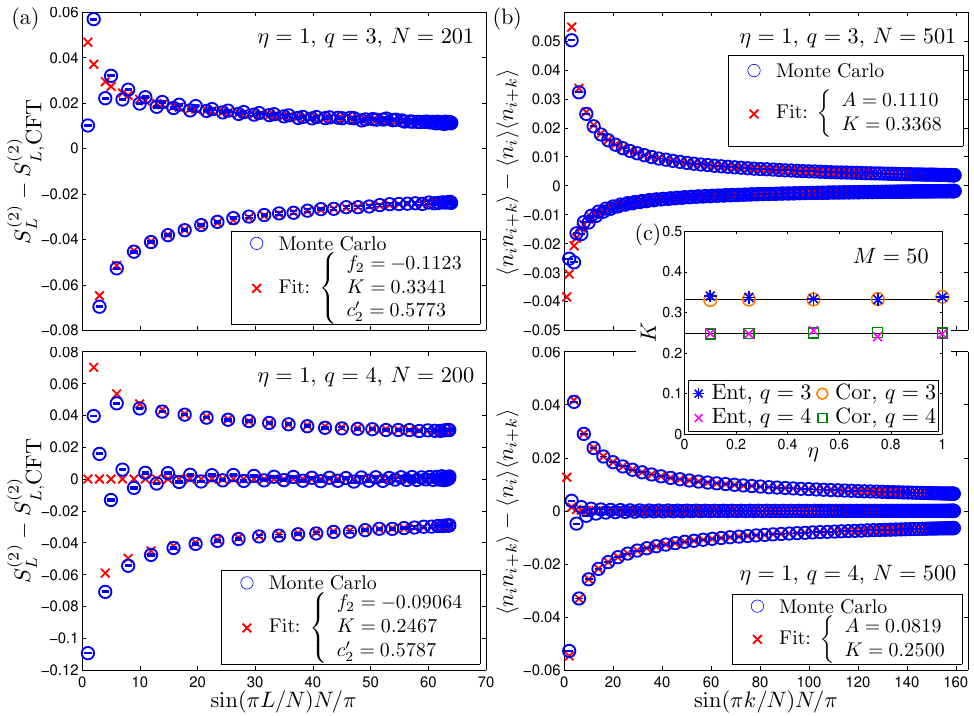}
\caption{(Color online) (a) Deviation of the R\'enyi entropy with index $\protect\alpha=2$ of a block of $L$ consecutive sites from the lowest order CFT expression \eqref{eq:SLCFT} and (b) particle-particle correlation function of the CFT state \eqref{eq:Laughlin} for a uniform 1D lattice in the lattice limit for $q=3$ (top) and $q=4$ (bottom) obtained from Monte Carlo simulations. The fits are based on Eqs.~\eqref{eq:SL} and \eqref{eq:cor}, respectively, and allow us to extract the Luttinger parameter $K$, which is shown for $M=50$ as a function of the interpolation parameter $\eta$ in inset (c) [`Ent' (`Cor') means extracted from the entropy (correlator) fit]. Since \eqref{eq:SL} and \eqref{eq:cor} are valid for large $L$ and $k$, respectively, we exclude the first $2q/\eta$ points when computing the fits.}
\label{fig:entcor}
\end{figure}

\textit{Continuous interpolation.--} We next demonstrate that important properties of the states \eqref{eq:Laughlin} stay the same as a function of the interpolation parameter, which indicates that the states remain within the same phase when interpolated between the lattice limit and the continuum limit. We first consider the uniform lattice in 1D and show that \eqref{eq:Laughlin} is well-described by the TLL theory in this case. The R\'enyi entropy $S_{L}^{(\alpha )}=\ln (\Tr(\rho _{L}^{\alpha }))/(1-\alpha )$ of a TLL, where $\rho _{L}$ is the reduced density operator of $L$ successive sites in the chain, is expected to be \cite{Calabrese-2010}
\begin{equation}
S_{L}^{(\alpha )}=S_{L,\text{CFT}}^{(\alpha )}+\frac{f_{\alpha }\cos (2Lk_\textrm{F})}{|2\sin (k_\textrm{F})\sin (\pi L/N)N/\pi |^{2K/\alpha }}  \label{eq:SL}
\end{equation}
for $\ln (|2\sin (k_\textrm{F})\sin (\pi L/N)N/\pi |)\gg \alpha $, where $K$ is the Luttinger parameter, $k_\textrm{F}=\eta\pi/q$ is the Fermi momentum,
\begin{equation}
S_{L,\text{CFT}}^{(\alpha )}=(c/6)(1+1/\alpha )\ln (\sin (\pi L/N)N/\pi
)+c_{\alpha }^{\prime },  \label{eq:SLCFT}
\end{equation}
$c$ is the central charge, and $f_{\alpha }$ and $c_{\alpha }^{\prime }$ are
nonuniversal constants. Fixing $c=1$ and using $f_{\alpha }$, $K$, and $c_{\alpha }^{\prime }$ as fitting parameters, we find that the entanglement entropy of (\ref{eq:Laughlin}), indeed, follows (\ref{eq:SL}) as illustrated for $\eta=1$ in Fig.~\ref{fig:entcor}(a). The expected TLL behavior of the particle-particle correlation function $C(k)=\langle n_{i}n_{i+k}\rangle-\langle n_{i}\rangle \langle n_{i+k}\rangle $ is \cite{Cabra-2004}
\begin{equation}
C(k)=\frac{A\cos (2kk_\textrm{F})}{|\sin (\pi k/N)N/\pi |^{2K}}+\frac{K}{2\pi
^{2}|\sin (\pi k/N)N/\pi |^{2}}  \label{eq:cor}
\end{equation}
for large $k$, where $A$ is a nonuniversal constant, and we find that this expression provides a good fit as illustrated for $\eta=1$ in Fig.~\ref{fig:entcor}(b). The values of $K$ extracted from the entropy and correlation function computations are shown as a function of the interpolation parameter in Fig.~\ref{fig:entcor}(c), and these results suggest that $K=1/q$ independent of $\eta$. We note that the observed behavior coincides with the properties of the free boson CFT with radius $R=\sqrt{q}$, which is the low-energy effective theory for the Calogero-Sutherland model with rational coupling constant $q$ \cite{Kawakami-1991}.

\begin{figure}[tbp]
\includegraphics[width=\columnwidth]{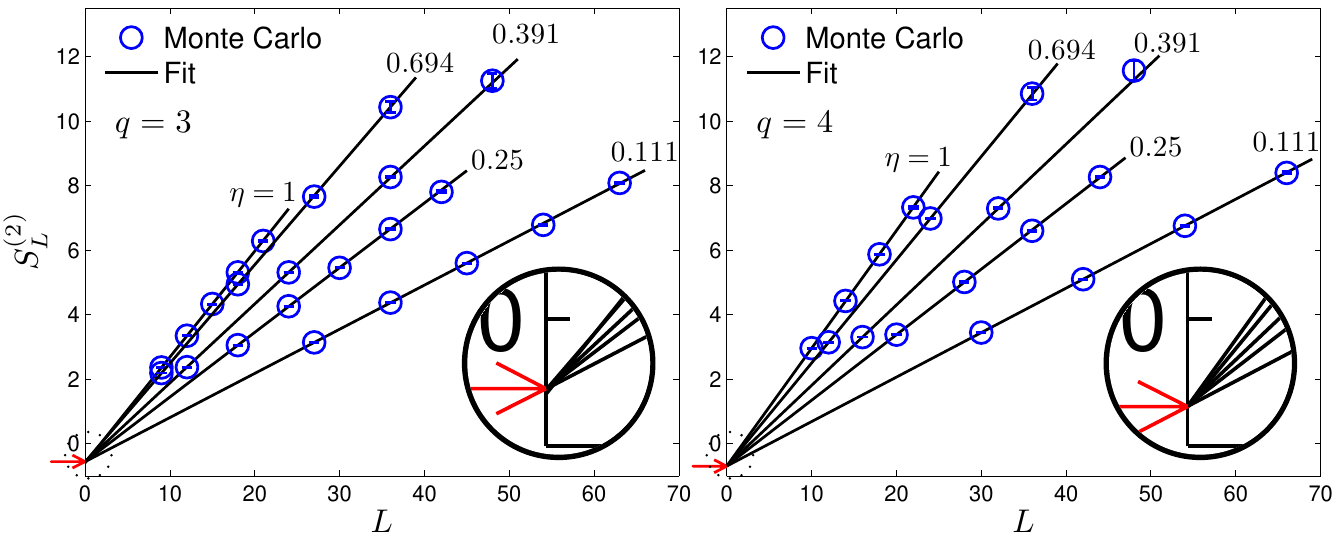}
\caption{(Color online) R\'{e}nyi entropy of the CFT state \eqref{eq:Laughlin} with $q=3$ (left) and $q=4$ (right) obtained from Monte Carlo simulations. The state is defined on an $R\times L$ square lattice on the cylinder, and the cut divides it into two $R/2\times L$ lattices, where $L$ is the number of lattice sites in the periodic direction. The fits are of the form $S_{L}^{(2)}=\xi L-\protect\gamma$, where $\xi$ and $\protect\gamma$ are fitting parameters, and are weighted so that points with larger error bars count less. Starting from above, $\protect\eta$ and $R$ are, respectively, $\protect\eta=1,0.694,0.391,0.25,0.111$ and $R=10,12,16,20,30$ for the five data sets, and the number of particles is $M=\eta RL/q$. The TEE $-\protect\gamma $ is seen to be independent of $\protect\eta $. The insets are enlarged views, and the red arrows point at the value $-\ln (q)/2$, which is the TEE of the Laughlin states in the continuum.}
\label{fig:tee}
\end{figure}

The Laughlin states in the continuum are topological states with TEE $-\ln(q)/2$, and in Fig.~\ref{fig:tee} we find that this value remains unchanged when interpolating the state to the lattice limit. The TEE $-\gamma $ is computed by mapping the state on an $R\times L$ square lattice to the cylinder, cutting the cylinder in two halves, computing the R\'{e}nyi entropy of one of the halves as a function of the number of sites $L$ along the cut, and utilizing that the R\'{e}nyi entropy follows the behavior $S_{L}^{(2)}=\xi L-\gamma $ for large $R$ and $L$, where $\xi$ is a nonuniversal constant \cite{Jiang-2012}. The mapping to the cylinder amounts to choosing $z_{j}=\exp (2\pi (r_{j}+il_{j})/L)$, where $r_{j}\in \{-R/2+1/2,-R/2+3/2,\ldots ,R/2-1/2\}$ and $l_{j}\in \{1,2,\ldots,L\}$. The CFT states in the lattice limit are therefore continuously connected to the Laughlin states in the continuum.

\textit{Parent Hamiltonian.--} For $\eta_j=1$ $\forall j$, the vertex operators constructing the wave functions (\ref{eq:iMPS}) can be identified as primary
fields of a free-boson CFT compactified on a circle of radius $R=\sqrt{q}$.
For $q=2$, the CFT is the SU(2)$_{1}$ Wess-Zumino-Witten (WZW) model. For $q=3$, the CFT has a hidden supersymmetry and can be identified as the $\mathcal{N}=2$ superconformal field theory \cite{Moore-Read-1991}. For integer $q$, the rationality of these CFTs ensures the existence of null fields. This is very useful, because null fields can be used for deriving parent Hamiltonians as demonstrated for the case of WZW models in \cite{nsc-2011}. Here, we identify a suitable set of null fields from which we derive decoupling equations. After some algebra \cite{SuppMat}, this procedure gives us a set of operators, which annihilate the wave functions (\ref{eq:iMPS}) at $\eta_j=1$. These operators include $\Upsilon =\sum_{i=1}^{N}\tilde{d}_{i}$, where $\tilde{d}_{i}=\chi_i^{-1}d_i$ and $d_i$ denotes the fermionic (hardcore bosonic) annihilation operator for odd (even) $q$, and
\begin{equation}
\Lambda _{i}=(q-2)\tilde{d}_{i}+\sum_{j(\neq i)}w_{ij}[\tilde{d}_{j}-\tilde{d}_{i}(qn_{j}-1)],
\label{eq:annihilator}
\end{equation}
where $w_{ij}\equiv (z_{i}+z_{j})/(z_{i}-z_{j})$. Since $\Upsilon|\Psi\rangle=\Lambda _{i}|\Psi\rangle=0$ $\forall i$, the positive semi-definite Hermitian operators $\Upsilon^\dagger\Upsilon$ and $\Lambda _{i}^{\dagger }\Lambda _{i}$ ($i=1,\ldots ,N$)\ have the wave functions (\ref{eq:iMPS}) with $\eta_j=1$ and $z_j$ arbitrary as their zero-energy ground state. Thus, these operators can be used to construct both 1D and 2D parent Hamiltonians for which the wave functions (\ref{eq:iMPS}) with $\eta_j=1$ are exact ground states. For the states with $\eta_j \neq 1$, we have not achieved to construct parent Hamiltonians, which is still an interesting open problem.

In the following, we focus on a 1D parent Hamiltonian obtained for $z_j=e^{2\pi ij/N}$, which turns out to have a particularly simple form. Specifically, we consider $H_{\mathrm{1D}}=\frac{1}{2}\sum_{i}(\Lambda_{i}^{\dagger }\Lambda _{i}-q\Gamma _{i}^{\dagger }\Gamma _{i})+\frac{q-2}{2}
\Upsilon ^{\dagger }\Upsilon +E_{0}$, where $\Gamma _{i}=\tilde{d}_{i}\Lambda _{i}=\sum_{j(\neq i)}w_{ij}\tilde{d}_{i}\tilde{d}_{j}$ and $E_{0}=-\frac{q-1}{6q}N[3N+(q-8)]$ is the eigenenergy of (\ref{eq:Laughlin}). This choice yields a parent Hamiltonian with purely two-body interactions
\begin{equation}
H_{\mathrm{1D}}=\sum_{i\neq j}
[(q-2)w_{ij}-w_{ij}^{2}]\tilde{d}_{i}^{\dagger }\tilde{d}_{j}-
\frac{q(q-1)}{2}\sum_{i\neq j}w_{ij}^{2}n_{i}n_{j}.
\label{eq:1DHamiltonian}
\end{equation}
While the $q=2$ Hamiltonian recovers the spin-1/2 Haldane-Shastry model \cite{Haldane-1988,Shastry-1988}, the Hamiltonians with $q\geq 3$ differ from the Haldane's inverse-square Hamiltonians \cite{Haldane-1988} by an extra hopping term. By diagonalizing the Hamiltonian (\ref{eq:1DHamiltonian}) numerically for small $N$, we confirm that the wave functions (\ref{eq:iMPS}) are indeed their unique ground states. Additionally, we observe that $H_{\mathrm{1D}}$ always has integer eigenvalues besides non-integer ones, which is an interesting feature already arising in Haldane's model \cite{Haldane-1988}. Motivated by Haldane's results, we have found that, after subtracting a constant, \textit{part of} the integer eigenvalues take the form $E=\sum_{\{m_{k}\}}2m_{k}(m_{k}+q-2-N)$, where $\{m_{k}\}$ is a set of $M$ pseudomomenta ($M$: number of particles) satisfying $m_{k}\in \lbrack 0,N-1]$ and $m_{k+1}\geq m_{k}+q$. This formula captures the essential low-lying part of the energy spectrum. Similar to Haldane's model, one can prove analytically that the Jastrow wave functions $\Psi _{\mathrm{1D}}^{J}(n_{1},\ldots,n_{N}) =\delta_n\prod_{i<j}(z_{i}-z_{j})^{qn_{i}n_{j}}\prod_{l}(\chi_lz_{l}^J)^{n_{l}}$, where $\delta_n=1$ for $\sum_in_i=M$ and zero otherwise and $1\leq J\leq N-q(M-1)-1$, are exact eigenstates of (\ref{eq:1DHamiltonian}) and are a subclass of those eigenstates with integer eigenvalues.

\textit{Conclusion.--} The present work combines several known models into a common framework with an underlying CFT structure and shows how the Laughlin states and the CS wave functions can be continuously transformed into lattice wave functions with similar properties. The CFT structure provides useful tools for deriving properties of the states analytically, and, in particular, enables us to derive parent Hamiltonians of the states in the lattice limit. Analytical wave functions play an important role in the investigation of the FQH effect in the continuum, and the model proposed here may similarly be used for analyzing FQH properties in lattice systems. Our present work also provides a method to discretize continuum FQH states in a way that is amenable to projected entangled-pair state description \cite{PEPS}, and thus it provides an alternative approach to the one recently introduced based on infinite matrix product states using discrete Landau level orbitals \cite{Zaletel-2012,Estienne-2013a,Estienne-2013b}.

\textit{Acknowledgment.--} We thank the Benasque Center of Sciences, where part of this work has been done, for their hospitality. This work has been supported by the EU project SIQS, the DFG cluster of excellence NIM, FIS2012-33642, QUITEMAD (CAM), and the Severo Ochoa Program.

\onecolumngrid
\appendix
\setcounter{equation}{0}
\newpage

\begin{center}
\textbf{Supplemental material}
\end{center}

\section{Operators annihilating the lattice Laughlin states}

In this section, we derive operators that annihilate the state (2) in the main text for $\eta=1$. We first assume $\chi_j=1$ and consider the CFT wave functions defined by%
\begin{equation}
\Psi _{n_{1},\ldots ,n_{N}}(z_{1},\ldots ,z_{N})=\langle
V_{n_{1}}(z_{1})V_{n_{2}}(z_{2})\cdots V_{n_{N}}(z_{N})\rangle ,
\label{eq:Laughlin}
\end{equation}%
where%
\begin{equation}
V_{n_{j}=1}(z_{j})=e^{i\pi (j-1)}V_{+}(z_{j})\text{ \ \ \ \ \ }%
V_{n_{j}=0}(z_{j})=V_{-}(z_{j}).
\end{equation}%
Here $V_{+}(z)=e^{i(q-1)\phi (z)/\sqrt{q}}$ and $V_{-}(z)=e^{-i\phi (z)/%
\sqrt{q}}$.

For the $c=1$ free-boson CFT with compactification radius $R=\sqrt{q}$, it
is convenient to define two chiral currents,%
\begin{equation}
G^{\pm }(z)=e^{\pm i\sqrt{q}\phi (z)},
\end{equation}%
besides the U(1) current $J(z)=\frac{i}{\sqrt{q}}\partial \phi (z)$. For $%
q=2 $, these currents form the SU(2)$_{1}$ Kac-Moody algebra. For $q=3$,
together with the energy-momentum tensor, the currents form the $\mathcal{N}%
=2$ superconformal current algebra.

To construct the parent Hamiltonian of (\ref{eq:Laughlin}), we need to
derive decoupling equations satisfied by the CFT correlator (\ref%
{eq:Laughlin}) using null fields. Let us first consider the null field%
\begin{eqnarray}
\chi _{1}(w) &=&\oint_{w}\frac{dz}{2\pi i}\frac{1}{z-w}%
[G^{+}(z)V_{-}(w)-qJ(z)V_{+}(w)]  \notag \\
&=&\oint_{w}\frac{dz}{2\pi i}\frac{1}{z-w}[e^{i\sqrt{q}\phi (z)}e^{-i\phi
(w)/\sqrt{q}}-\sqrt{q}i\partial \phi (z)e^{i(q-1)\phi (w)/\sqrt{q}}]  \notag
\\
&=&\oint_{w}\frac{dz}{2\pi i}\frac{1}{z-w}[\frac{1}{z-w}e^{i\sqrt{q}\phi
(z)-i\phi (w)/\sqrt{q}}-\sqrt{q}i\partial \phi (w)e^{i(q-1)\phi (w)/\sqrt{q}%
}]  \notag \\
&=&\oint_{w}\frac{dz}{2\pi i}\frac{1}{z-w}[\sqrt{q}i\partial \phi
(w)e^{i(q-1)\phi (w)/\sqrt{q}}-\sqrt{q}i\partial \phi (w)e^{i(q-1)\phi (w)/%
\sqrt{q}}]  \notag \\
&=&0.
\end{eqnarray}%
By replacing the vertex operator at site $i$ by the null field $\chi
_{1}(z_{i})$, the chiral correlator vanishes%
\begin{eqnarray*}
0 &=&\langle V_{n_{1}}(z_{1})\cdots \chi _{1}(z_{i})\cdots
V_{n_{N}}(z_{N})\rangle  \\
&=&\oint_{z_{i}}\frac{dz}{2\pi i}\frac{1}{z-z_{i}}\langle
V_{n_{1}}(z_{1})\cdots \lbrack G^{+}(z)V_{-}(z_{i})-qJ(z)V_{+}(z_{i})]\cdots
V_{n_{N}}(z_{N})\rangle  \\
&=&-\sum_{j=1(\neq i)}^{N}\oint_{z_{j}}\frac{dz}{2\pi i}\frac{1}{z-z_{i}}%
\langle V_{n_{1}}(z_{1})\cdots \lbrack
G^{+}(z)V_{-}(z_{i})-qJ(z)V_{+}(z_{i})]\cdots V_{n_{N}}(z_{N})\rangle ,
\end{eqnarray*}%
where we have deformed the integral contour in the last step. To proceed we
use the operator product expansions (OPEs)%
\begin{eqnarray}
G^{+}(z)V_{n}(w) &\sim &\frac{\sum_{n^{\prime }}(d)_{nn^{\prime }}}{z-w}%
V_{n^{\prime }}(w), \\
J(z)V_{n}(w) &\sim &\frac{1}{q}\frac{\sum_{n^{\prime }}(qd^{\dagger
}d-1)_{nn^{\prime }}}{z-w}V_{n^{\prime }}(w),
\end{eqnarray}%
where the particle annihilation and creation operators are defined as $d=%
\begin{pmatrix}
0 & 0 \\
1 & 0%
\end{pmatrix}%
$ and $d^{\dagger }=%
\begin{pmatrix}
0 & 1 \\
0 & 0%
\end{pmatrix}%
$, respectively. Applying the OPEs, the chiral correlator with null field $%
\chi _{1}(z_{i})$ yields the following decoupling equation:%
\begin{eqnarray}
0 &=&\langle V_{n_{1}}(z_{1})\cdots \chi _{1}(z_{i})\cdots
V_{n_{N}}(z_{N})\rangle   \notag \\
&=&-\sum_{j=1(\neq i)}^{N}\oint_{z_{j}}\frac{dz}{2\pi i}\frac{1}{z-z_{i}}%
\langle V_{n_{1}}(z_{1})\cdots \lbrack
G^{+}(z)V_{-}(z_{i})-qJ(z)V_{+}(z_{i})]\cdots V_{n_{N}}(z_{N})\rangle
\notag \\
&=&-\sum_{j=1(\neq i)}^{N}\oint_{z_{j}}\frac{dz}{2\pi i}\frac{1}{z-z_{i}}%
\frac{\sum_{n_{j}^{\prime }}(d)_{n_{j}n_{j}^{\prime }}}{z-z_{j}}\langle
V_{n_{1}}(z_{1})\cdots V_{n_{j}^{\prime }}(z_{j})\cdots V_{-}(z_{i})\cdots
V_{n_{N}}(z_{N})\rangle   \notag \\
&&+\sum_{j=1(\neq i)}^{N}\oint_{z_{j}}\frac{dz}{2\pi i}\frac{1}{z-z_{i}}%
\frac{\sum_{n_{j}^{\prime }}(qd^{\dagger }d-1)_{n_{j}n_{j}^{\prime }}}{%
z-z_{j}}\langle V_{n_{1}}(z_{1})\cdots V_{n_{j}^{\prime }}(z_{j})\cdots
V_{+}(z_{i})\cdots V_{n_{N}}(z_{N})\rangle   \notag \\
&=&\sum_{j=1(\neq i)}^{N}\frac{1}{z_{i}-z_{j}}\sum_{n_{j}^{\prime
}}(d)_{n_{j}n_{j}^{\prime }}\langle V_{n_{1}}(z_{1})\cdots V_{n_{j}^{\prime
}}(z_{j})\cdots V_{-}(z_{i})\cdots V_{n_{N}}(z_{N})\rangle   \notag \\
&&-\sum_{j=1(\neq i)}^{N}\frac{1}{z_{i}-z_{j}}\sum_{n_{j}^{\prime
}}(qd^{\dagger }d-1)_{n_{j}n_{j}^{\prime }}\langle V_{n_{1}}(z_{1})\cdots
V_{n_{j}^{\prime }}(z_{j})\cdots V_{+}(z_{i})\cdots V_{n_{N}}(z_{N})\rangle .
\end{eqnarray}%
Based on the above decoupling equation, we obtain an operator $\Lambda
_{i}^{\prime }$%
\begin{equation}
\Lambda _{i}^{\prime }=\sum_{j=1(\neq i)}^{N}\frac{1}{z_{i}-z_{j}}%
[d_{i}^{\dagger }d_{j}-n_{i}(qn_{j}-1)],
\end{equation}%
where $n_{j}=d_{j}^{\dagger }d_{j}$, and which annihilates the wave function
(\ref{eq:Laughlin}), i.e., $\Lambda _{i}^{\prime }|\Psi \rangle =0$ $\forall
i=1,\ldots ,N$.

Similarly, decoupling equations can be derived from another two null fields%
\begin{eqnarray}
\chi _{2}(w) &=&\oint_{w}\frac{dz}{2\pi i}\frac{1}{z-w}G^{+}(z)V_{+}(w)=0, \\
\chi _{3}(w) &=&\oint_{w}\frac{dz}{2\pi i}G^{+}(z)V_{+}(w)=0,
\end{eqnarray}%
and we obtain two additional operators annihilating the wave function (\ref%
{eq:Laughlin})%
\begin{eqnarray}
\Lambda _{i}^{\prime \prime } &=&\sum_{j=1(\neq i)}^{N}\frac{1}{z_{i}-z_{j}}%
n_{i}d_{j}, \\
\Upsilon &=&\sum_{i=1}^{N}d_{i}.
\end{eqnarray}%
These operators can be combined into new operators annihilating (\ref%
{eq:Laughlin})
\begin{eqnarray}
d_{i}\Lambda _{i}^{\prime }+\Lambda _{i}^{\prime \prime } &=&\sum_{j=1(\neq
i)}^{N}\frac{1}{z_{i}-z_{j}}[d_{j}-d_{i}(qn_{j}-1)], \\
d_{i}\Lambda _{i}^{\prime \prime } &=&\sum_{j=1(\neq i)}^{N}\frac{1}{%
z_{i}-z_{j}}d_{i}d_{j}.
\end{eqnarray}

Defining $w_{ij}=\frac{z_{i}+z_{j}}{z_{i}-z_{j}}$, the operator $\Lambda
_{i}=(q-2)d_{i}+\sum_{j=1(\neq i)}^{N}w_{ij}[d_{j}-d_{i}(qn_{j}-1)]$ can be
written as%
\begin{eqnarray*}
\Lambda _{i} &=&(q-2)d_{i}+\sum_{j=1(\neq i)}^{N}\left( \frac{2z_{i}}{%
z_{i}-z_{j}}-1\right) [d_{j}-d_{i}(qn_{j}-1)] \\
&=&(q-2)d_{i}+2z_{i}(d_{i}\Lambda _{i}^{\prime }+\Lambda _{i}^{\prime \prime
})-\sum_{j=1(\neq i)}^{N}[d_{j}-d_{i}(qn_{j}-1)] \\
&=&(q-2)d_{i}+2z_{i}(d_{i}\Lambda _{i}^{\prime }+\Lambda _{i}^{\prime \prime
})-(\Upsilon -d_{i})+d_{i}\left[ \sum_{j=1}^{N}(qn_{j}-1)-(qn_{i}-1)\right]
\\
&=&2z_{i}(d_{i}\Lambda _{i}^{\prime }+\Lambda _{i}^{\prime \prime
})-\Upsilon +d_{i}\sum_{j=1}^{N}(qn_{j}-1).
\end{eqnarray*}%
Note that the wave function (\ref{eq:Laughlin}) has filling fraction $\nu
=1/q$, i.e., $\sum_{j=1}^{N}(qn_{j}-1)|\Psi \rangle =0$. Thus, we have
proven that $\Lambda _{i}|\Psi \rangle =0$ $\forall i=1,\ldots ,N$. Since $\Lambda _{i}|\Psi \rangle =0$, it is straightforward to prove that $%
\Gamma _{i}|\Psi \rangle =0$, where $\Gamma _{i}$ is given by $\Gamma
_{i}=d_{i}\Lambda _{i}=\sum_{j=1(\neq i)}^{N}w_{ij}d_{i}d_{j}$.

The wave function in the main text for $\eta=1$ differs from \eqref{eq:Laughlin} by the factor $\prod_j\chi_j^{n_j}$. This can, however, easily be taken into account by multiplying the above operators with $\prod_j\chi_j^{-n_j}$ from the right and $\prod_j\chi_j^{n_j}$ from the left, which amounts to replacing $d_j$ by $\tilde{d}_j=\chi_j^{-1}d_j$.

\section{1D parent Hamiltonian}

In this section, we use $\Lambda _{i}$ to construct a 1D uniform
Hamiltonian, where the lattice sites form a unit circle, i.e., $%
z_{j}=e^{i2\pi j/N}$.

Since $\sum_{j(\neq i)}w_{ij}=0$ in 1D uniform case, the form of $\Lambda
_{i}$ can be simplified as
\begin{equation*}
\Lambda _{i}=(q-2)d_{i}+\sum_{j(\neq i)}w_{ij}(d_{j}-qd_{i}n_{j}).
\end{equation*}%
Then, the positive-semidefinite operators annihilating the wave functions
are given by%
\begin{eqnarray*}
\Lambda _{i}^{\dagger }\Lambda _{i} &=&(q-2)^{2}d_{i}^{\dagger
}d_{i}+(q-2)\sum_{j(\neq i)}w_{ij}(d_{i}^{\dagger
}d_{j}-qn_{i}n_{j})-(q-2)\sum_{j(\neq i)}w_{ij}(d_{j}^{\dagger
}d_{i}-qn_{i}n_{j}) \\
&&-\sum_{j(\neq i)}w_{ij}^{2}(d_{j}^{\dagger }-qd_{i}^{\dagger
}n_{j})(d_{j}-qd_{i}n_{j})-\sum_{j\neq l(\neq i)}w_{ij}w_{il}(d_{l}^{\dagger
}-qd_{i}^{\dagger }n_{l})(d_{j}-qd_{i}n_{j}) \\
&=&(q-2)^{2}n_{i}+(q-2)\sum_{j(\neq i)}w_{ij}(d_{i}^{\dagger
}d_{j}-d_{j}^{\dagger }d_{i}) \\
&&-\sum_{j(\neq i)}w_{ij}^{2}(n_{j}+q^{2}n_{i}n_{j})-\sum_{j\neq l(\neq
i)}w_{ij}w_{il}[d_{l}^{\dagger }d_{j}-q(d_{j}^{\dagger }d_{i}+d_{i}^{\dagger
}d_{j})n_{l}+q^{2}n_{i}n_{j}n_{l}].
\end{eqnarray*}%
By using the useful identities%
\begin{eqnarray*}
\sum_{i(\neq j)}w_{ij}^{2} &=&-\frac{(N-1)(N-2)}{3}, \\
\sum_{i(\neq j,l)}w_{ij}w_{il} &=&(N-2)+2w_{jl}^{2},
\end{eqnarray*}%
and fixing the filling fraction $\sum_{i}n_{i}=N/q$ in the system, we obtain%
\begin{eqnarray*}
\sum_{i}\Lambda _{i}^{\dagger }\Lambda _{i} &=&(q-2)^{2}\frac{N}{q}%
+2(q-2)\sum_{i\neq j}w_{ij}d_{i}^{\dagger }d_{j}+\frac{(N-1)(N-2)}{3}%
\sum_{j}n_{j}-q^{2}\sum_{i\neq j}w_{ij}^{2}n_{i}n_{j} \\
&&-\sum_{j\neq l}[(N-2)+2w_{jl}^{2}]d_{l}^{\dagger }d_{j}+q\sum_{i\neq j\neq
l}w_{ij}w_{il}[(d_{j}^{\dagger }d_{i}+d_{i}^{\dagger
}d_{j})n_{l}-qn_{i}n_{j}n_{l}] \\
&=&(q-2)^{2}\frac{N}{q}+\frac{N(N-1)(N-2)}{3q}+2(q-2)\sum_{i\neq
j}w_{ij}d_{i}^{\dagger }d_{j}-q^{2}\sum_{i\neq j}w_{ij}^{2}n_{i}n_{j} \\
&&-\sum_{j\neq l}[(N-2)+2w_{jl}^{2}]d_{l}^{\dagger }d_{j}+q\sum_{i\neq j\neq
l}w_{ij}w_{il}[(d_{j}^{\dagger }d_{i}+d_{i}^{\dagger
}d_{j})n_{l}-qn_{i}n_{j}n_{l}].
\end{eqnarray*}

The above expression can be further simplified by using%
\begin{equation*}
\sum_{j\neq l}d_{l}^{\dagger }d_{j}=\Upsilon ^{\dagger }\Upsilon -\frac{N}{q}
\end{equation*}%
and%
\begin{eqnarray*}
\sum_{i\neq j\neq l}w_{ij}w_{il}n_{i}n_{j}n_{l} &=&\frac{1}{3}\sum_{i\neq
j\neq l}(w_{ij}w_{il}+w_{ji}w_{jl}+w_{li}w_{lj})n_{i}n_{j}n_{l} \\
&=&\frac{1}{3}\sum_{i\neq j\neq l}n_{i}n_{j}n_{l} \\
&=&\frac{N(N-q)(N-2q)}{3q^{3}},
\end{eqnarray*}%
where we have used the cyclic identity%
\begin{equation*}
w_{ij}w_{il}+w_{ji}w_{jl}+w_{li}w_{lj}=1.
\end{equation*}%
Then, we obtain%
\begin{eqnarray}
\sum_{i}\Lambda _{i}^{\dagger }\Lambda _{i} &=&(q-2)^{2}\frac{N}{q}+\frac{%
N(N-1)(N-2)}{3q}+2(q-2)\sum_{i\neq j}w_{ij}d_{i}^{\dagger
}d_{j}-q^{2}\sum_{i\neq j}w_{ij}^{2}n_{i}n_{j}  \notag \\
&&-(N-2)(\Upsilon ^{\dagger }\Upsilon -\frac{N}{q})-2\sum_{i\neq
j}w_{ij}^{2}d_{i}^{\dagger }d_{j}+q\sum_{i\neq j\neq
l}w_{ij}w_{il}(d_{j}^{\dagger }d_{i}+d_{i}^{\dagger }d_{j})n_{l}-q^{2}\frac{%
N(N-q)(N-2q)}{3q^{3}}  \notag \\
&=&2\sum_{i\neq j}[(q-2)w_{ij}-w_{ij}^{2}]d_{i}^{\dagger
}d_{j}-q^{2}\sum_{i\neq j}w_{ij}^{2}n_{i}n_{j}+q\sum_{i\neq j\neq
l}w_{ij}w_{il}(d_{j}^{\dagger }d_{i}+d_{i}^{\dagger
}d_{j})n_{l}-(N-2)\Upsilon ^{\dagger }\Upsilon  \notag \\
&&+\frac{N}{3q}[3qN+(q^{2}-12q+8)].
\end{eqnarray}

Now we construct positive-semidefinite operators from the operator $\Gamma
_{i}=\sum_{j(\neq i)}w_{ij}d_{i}d_{j}$%
\begin{eqnarray*}
\Gamma _{i}^{\dagger }\Gamma _{i} &=&-\sum_{j,l(\neq
i)}w_{ij}w_{il}d_{l}^{\dagger }d_{j}n_{i} \\
&=&-\sum_{j(\neq i)}w_{ij}^{2}n_{i}n_{j}-\sum_{j\neq l(\neq
i)}w_{ij}w_{il}d_{l}^{\dagger }d_{j}n_{i},
\end{eqnarray*}%
and%
\begin{equation}
\sum_{i}\Gamma _{i}^{\dagger }\Gamma _{i}=-\sum_{i\neq
j}w_{ij}^{2}n_{i}n_{j}-\sum_{i\neq j\neq l}w_{lj}w_{li}d_{i}^{\dagger
}d_{j}n_{l}.
\end{equation}

Note that $\sum_{i}\Lambda _{i}^{\dagger }\Lambda _{i}$ and $\sum_{i}\Gamma
_{i}^{\dagger }\Gamma _{i}$ both contain three-body interaction terms.
However, we observe that, the following combination eliminates the
three-body terms by using the cyclic identity:%
\begin{eqnarray*}
&&\sum_{i}\Lambda _{i}^{\dagger }\Lambda _{i}-q\sum_{i}\Gamma _{i}^{\dagger
}\Gamma _{i} \\
&=&2\sum_{i\neq j}[(q-2)w_{ij}-w_{ij}^{2}]d_{i}^{\dagger
}d_{j}-(q^{2}-q)\sum_{i\neq j}w_{ij}^{2}n_{i}n_{j}-(N-2)\Upsilon ^{\dagger
}\Upsilon \\
&&+q\sum_{i\neq j\neq
l}(w_{ij}w_{il}+w_{ji}w_{jl}+w_{lj}w_{li})d_{i}^{\dagger }d_{j}n_{l}+\frac{N%
}{3q}[3qN+(q^{2}-12q+8)] \\
&=&2\sum_{i\neq j}[(q-2)w_{ij}-w_{ij}^{2}]d_{i}^{\dagger
}d_{j}-(q^{2}-q)\sum_{i\neq j}w_{ij}^{2}n_{i}n_{j}-(N-2)\Upsilon ^{\dagger
}\Upsilon \\
&&+q\sum_{i\neq j\neq l}d_{i}^{\dagger }d_{j}n_{l}+\frac{N}{3q}%
[3qN+(q^{2}-12q+8)] \\
&=&2\sum_{i\neq j}[(q-2)w_{ij}-w_{ij}^{2}]d_{i}^{\dagger
}d_{j}-(q^{2}-q)\sum_{i\neq j}w_{ij}^{2}n_{i}n_{j}-(q-2)\Upsilon ^{\dagger
}\Upsilon +\frac{q-1}{3q}N[3N+(q-8)],
\end{eqnarray*}%
where we have used%
\begin{eqnarray*}
\sum_{i\neq j\neq l}d_{i}^{\dagger }d_{j}n_{l} &=&\sum_{i\neq
j}d_{i}^{\dagger }d_{j}(\frac{N}{q}-n_{i}-n_{j}) \\
&=&(\frac{N}{q}-1)\sum_{i\neq j}d_{i}^{\dagger }d_{j} \\
&=&(\frac{N}{q}-1)\Upsilon ^{\dagger }\Upsilon -\frac{N}{q}(\frac{N}{q}-1).
\end{eqnarray*}

Finally, we define the 1D parent Hamiltonian as%
\begin{eqnarray}
H_{\mathrm{1D}} &=&\frac{1}{2}\sum_{i}\Lambda _{i}^{\dagger }\Lambda _{i}-%
\frac{q}{2}\sum_{i}\Gamma _{i}^{\dagger }\Gamma _{i}+\frac{q-2}{2}\Upsilon
^{\dagger }\Upsilon -E_{0}  \notag \\
&=&\sum_{i\neq j}[(q-2)w_{ij}-w_{ij}^{2}]d_{i}^{\dagger }d_{j}-\frac{1}{2}%
(q^{2}-q)\sum_{i\neq j}w_{ij}^{2}n_{i}n_{j},
\end{eqnarray}%
where $E_{0}$ is the ground-state energy of $H_{\mathrm{1D}}$
\begin{equation}
E_{0}=-\frac{q-1}{6q}N[3N+(q-8)].
\end{equation}
If $\chi_j\neq1$, $d_j$ should be replaced by $\tilde{d}_j=\chi_j^{-1}d_j$.

\end{document}